\documentclass[12pt]{article}

\usepackage{graphicx}
\usepackage{amssymb}
\usepackage{amsthm}
\usepackage{amsmath}
\usepackage{hyperref}
\usepackage{bm}
\usepackage{authblk}

\newcommand{\pvalue}{\emph{p}-value}

\title{Percent Change Estimation in Large Scale Online Experiments}
\author{Jacopo Soriano}
\affil{Google Inc.}

\begin{document}
\maketitle

\begin{abstract}
Online experiments are a fundamental component of the development of web-facing products. Given their large user-bases, even small product improvements can have a large impact on user engagement or profits on an absolute scale. As a result, accurately estimating the relative impact of these changes is extremely important. I propose an approach based on an objective Bayesian model to improve the sensitivity of percent change estimation in A/B experiments. Leveraging pre-period information, this approach produces more robust and accurate point estimates and up to 50\% tighter credible intervals than traditional methods. The R package \texttt{abpackage} provides an implementation of the approach.
\end{abstract}

\section{Introduction}

Tech companies like Amazon, Facebook, Google and Microsoft rely on A/B experiments to evaluate the impact of potential product changes. Examples of product changes can range from new features in the user interface to algorithmic variations of a recommendation system. 
In an A/B experiment a small subset of users is selected to be in the experiment, with each user randomly assigned to the current state of the product (control group or group A) or to the potential change to the product (treatment group or group B). The experiment can last from a few days to several months depending on the nature of the treatment. Once the experiment is completed, the experimenter is generally interested in testing whether the average treatment effect is non-zero on key metrics like daily active users (DAUs), time on site, latency, revenue, etc.

The most common and simplest statistical procedure for hypothesis testing in this context is the t-test. However, despite the large number of users in these experiments, the t-test often lacks the statistical power necessary to detect tiny changes. This can be an issue when even a 0.1\% change can represent millions of dollars per year. To combat this issue, in recent years variance reduction techniques have been successfully developed and applied at tech companies like Microsoft \cite{deng2013improving} and Netflix \cite{xie2016improving}  to improve the statistical power of these tests. The key idea of these techniques is to incorporate 
pre-treatment covariates into the analysis to reduce variance. In most cases the most effective pre-treatment covariate is the metric of interest computed during a fixed time window prior to the start of the experiment.
 
Beyond hypothesis testing, it is important to estimate the size of the effect of the treatment. A good measure of the effect size is the percent change between the mean of the metric for the treatment group and the mean of the metric for the control group. Unlike the mean difference between the control and the treatment group, this quantity is scale-free. Having a scale free parameter is attractive because it facilitates comparisons across different experiments and different metrics. Additionally, the percent change is interpretable. If one scales this experiment to 100\% of traffic, for example, one expects an increase of x\% in the metric of interest.
 
Percent change estimation has been largely discussed in the medical literature  \cite{kaiser1989adjusting} and \cite{vickers2001use}, but it has not been carefully studied in the context of large scale online experiments. To my knowledge, the only papers discussing percent change estimation for online experimentation are \cite{kohavi2009controlled} and \cite{chamandy2012estimating}. However, in each of these papers percent change is mentioned only as an extension of the primary methodology being discussed, and variance reduction techniques are not considered. 

In the medical context the percent change is generally computed between the post-period and the pre-period in the same patient. Unfortunately, in online experiments there is often large day-to-day variability in the metrics of interest. As a result, the percent change between the pre-period and the post-period is generally not meaningful, and instead the focus is on the percent change between treatment and control in the post-period. 
Even if there is no direct interest in the pre-period, \cite{deng2013improving} showed that the pre-period can be incorporated into the analysis as a baseline covariate to reduce variance in the post-period.   
 
In this paper I propose a sensitive statistical method for percent change estimation in A/B experiments that effectively leverages the pre-period information.
Specifically, this new  approach is based on a two-stage objective Bayesian model to estimate the percent change in the post-period while controlling for a single pre-period covariate. The resulting point estimates are substantially more precise, and the credible intervals (CIs) are up to 50\% tighter than CIs based on traditional methods that do not correct for pre-period covariates.
 
Compared to classical methods, the advantages of using a Bayesian approach are two-fold. First, propagation of uncertainty across the two stages of the model is very natural in a Bayesian set up. Secondly, inference on an arbitrary function of the parameters is also straightforward. In this case, the function is the percent change between the two means. In addition, by using objective priors, the experimenter has a statistical procedure with good frequentist properties without having to elicit prior parameters. This allows the method to easily generalize to a variety of experiments and metrics.

This paper is organized as follows. Section~\ref{sec:bac} provides some background on existing approaches. Specifically,  
classical approaches for inference on  percent change and variance reduction techniques for hypothesis testing are described. In Section~\ref{sec:met}, 
I introduce a new method that combines percent change estimation and variance reduction. An algorithm to speed up inference is also described. In Section~\ref{sec:emp}, 
the methodology is iullstrated on several real and simulated data examples. In Section~\ref{sec:con}, the main contributions of this work are summarized.

Code to use this methodology are freely available at \url{https://google.github.io/abpackage} in the form of an R package called \texttt{abpackage}. 

\section{Background}\label{sec:bac}

Assume that the values of the metric of interest for the treatment and the control groups are random variables $Y_{i,j}$ where $i = 1, \ldots, n_j$ denotes observations in group j, and  $j=c, t$ denotes the control and the treatment groups, respectively. In large online experiments, data are often randomly \emph{bucketed}, i.e, data are the result of an aggregation of the metric across many users \cite{chamandy2012estimating}. This aggregation is done to avoid computational roadblocks, to reduce storage cost, and due to privacy considerations. For the DAUs metric, for example, the observation $Y_{i,t}$ represents the sum of the number of days that users that are in \emph{bucket} $i$ and in the treatment group used the product across the $T$ days of the experiment. Since observations are the result of an aggregation across users and days, by the Central Limit Theorem it is reasonable to assume that the underlying random variables are normally distributed
\begin{equation}\label{eq:post}
Y_{i,j} \sim N(\mu_j, \sigma_j^2).
 \end{equation}

 For strictly positive metrics, a good measure of effect size is the percent change between the treatment mean and the control mean
 \begin{equation}\label{eq:per}
100 \cdot (\mu_t - \mu_c) / \mu_c = 100 \cdot \mu_t / \mu_c - 100.
 \end{equation}
As described earlier, this quantity is attractive because it is scale free, allowing natural comparison across several metrics within the same experiment or across experiments. One caveat is that the CI for \eqref{eq:per} is well defined only when the estimate of the control mean $\mu_c$ is far from zero. This condition will be met whenever the sample mean $\bar{y}_c$ is sufficiently larger than the standard error $ SE_{\bar{y}_c}  = \sqrt{s_c^2 / n_c}$, where $s_c^2$ is the sample variance of the control group. A simple rule of thumb that I would suggest is $\bar{y}_c > 5 \cdot SE_{\bar{y}_c} $. In practice, this is generally not an issue.

\subsection{Classical Approaches for Percent Change Estimation}\label{sec:classic}

Several approaches to estimate \eqref{eq:per} have been proposed in the literature. The most common approaches include the  bootstrap \cite{efron1992bootstrap},  the Taylor's method, the Fieller's method \cite{fieller1940biological} and the Index method  \cite{franz2007ratios}. All these methods use the same point estimate $\bar{y}_t  / \bar{y}_c$, but they differ in their  estimation of the standard error for the percent change. In the remainder of this section I briefly describe how each of these approaches estimate the standard error. For a more in depth review see \cite{franz2007ratios}. 

The Taylor's method relies on a Taylor's expansion of the distribution of $f(\bar{Y}_t, \bar{Y}_c)$, where 
$\bar{Y}_j = \sum_i Y_{i,j} / n_j$ for $j=c,t$ and 
$f(x, z) = 100 \cdot x / z - 100$.  Using a first order expansion, the resulting standard error is equal to
 \begin{equation}\label{eq:taylor}
\dfrac{\mu_t}{\mu_c} \sqrt{\dfrac{\sigma_t^2}{\mu_t^2} + \dfrac{\sigma_c^2}{\mu_c^2}}.
 \end{equation}
An estimate of the  standard error can be obtained by using the plug-in principle, i.e., replacing the parameters $\mu_t$, $\mu_c$, $\sigma_c^2$ and $\sigma_t^2$ with their estimates $\bar{y}_t$, $\bar{y}_c$, $s_c^2$ and $s_t^2$.

The bootstrap is based on sampling with replacement observations in the treatment group and in the control group, respectively. For each of these samples $\bm{y}_t^{(s)}$ and $\bm{y}_c^{(s)}$ for $s= 1, \ldots, S$, the percent change $\bar{r}^{(s)} := 100 \cdot \bar{y}_t^{(s)} / \bar{y}_c^{(s)} - 100$ is computed. $S$ is the number of bootstrap samples, which is generally set to 5,000 or 10,000. Finally, the standard error can be obtained by computing the standard deviation of the bootstrap estimates $\bar{r}^{(s)}$  for $s= 1, \ldots, S$. 

The Fieller's method relies on the observation that if $\bar{Y}_t$ and $\bar{Y}_c$ are normally distributed, then $\bar{Y}_t - \mu_t / \mu_c \cdot \bar{Y}_c$ is also normally distributed. 

The Index method is based on pairing each observation in the treatment group with an observation in the control group, i.e., $r_i := 100 \cdot y_{i, t} / y_{i, c} - 100$.  The standard error is computed by calculating the t-test standard error of  these transformed observations $r_i$ for $i = 1, \ldots, n$. This approach relies on the assumption that the sample size of the control is equal to the sample size of the treatment, i.e., $n_t = n_c = n$. This approach can be inefficient when $n$ is small, since the degrees of freedom to estimate the standard error are cut by half by pairing each observation in the treatment with an observation in the control.

\subsection{Variance Reduction}\label{sub:var}

Online experiments can include million of users. Despite this, the statistical power of the classical approaches  for percent change estimation discussed in Section \ref{sec:classic}  can be insufficient because the signal to noise ratio is often very small. To overcome this issue \cite{deng2013improving} used a variance reduction technique called control-variate. 

The idea of control-variate is to identify a baseline metric $X$ that has the same distribution for the treatment and the control group such that $|Cor(X_{i,j}, Y_{i,j})| = |\rho_j| > 0$ for $j=t,c$.
 
Specifically, if $(X_{i,j}, Y_{i,j})$ is Normally distributed, and $X_{i,j}$ has mean $\mu_0$ and variance $\sigma_0^2$, the conditional distribution has the following expression
\begin{equation}\label{eq:cond}
Y_{i,j} | X_{i,j} = x_{i,j} \sim N \big( \mu_j + \beta_j \cdot (x_{i,j} - \mu_0), \tau_j^2 \big), 
\end{equation}
where $i = 1, \ldots, n_j$, $j = c, t$, $\beta_j := \rho_j \cdot \sigma_j / \sigma_0$ and $\tau_j^2 := (1 - \rho_j^2) \cdot \sigma_j^2$. 

Assume, for instance, that the correlation $\rho_j > 0$. Then, the expected mean for $Y_{i,j}$ is higher (lower) than the post-period mean $\mu_j$ when $x_{i,j}$ is higher (lower) than the baseline mean $\mu_0$.  In other words, $x_{i,j}$ is predictive of what one will observe in the post-period. As a result, the baseline metric $X$ removes some of the variability of $Y$. Specifically, the variance of \eqref{eq:cond} is reduced by a factor $1 - \rho_j^2$ with respect to the variance of \eqref{eq:post}. 
I.e., the magnitude of the variance reduction scales with the magnitude of the correlation.  When the two correlations are null, the distributions have the same variance.

In general, a high correlation can be achieved by setting $X$ equal to the  same metric as $Y$, but computed prior to the start of the experiment. 
For example, for the DAUs metric, $X_{i,j}$ would be the sum of the number of days that users in bucket $i$ and in the condition group $j$ used the product in the $T'$ days preceding the start of the experiment. 
From now on, I will refer to $X$ as the pre-period metric. 

In online experiments at YouTube it is not uncommon to observe correlations between the pre-period and the post-period of the order of 0.8 for both the control and the treatment, which results in variance reduction of over 30\%. 
The length of the pre-period may have an impact on the correlation. In practice I have observed that there are minimal incremental gains when considering a pre-period longer than a week, which is consistent with the findings in \cite{deng2013improving}.

The model proposed by \cite{deng2013improving} is a special case of \eqref{eq:cond}. Specifically, their model relies on the following regression,
$$
Y_{i,j} | X_{i,j} = x_{i,j} \sim N \big( \tilde{\mu}_j + \tilde{\beta} \cdot x_{i,j}, \tilde{\tau}^2 \big), 
$$
where $i = 1, \ldots, n_j$ and $j=t,c$, $\tilde{\mu}_j $ is the counterpart of $\mu_j + \beta_j \cdot \mu_0$ in \eqref{eq:cond}, $\tilde{\beta}$ is the counterpart of $\beta_j$  in \eqref{eq:cond} and $\tilde{\tau}^2$ is the counterpart of $\tau_j^2$  in \eqref{eq:cond}.
\cite{deng2013improving} rely on this parametrization of the linear predictor for two reasons. First, the means  $\mu_t, \mu_c$ and $\mu_0$ are not jointly identifiable in a frequentist model. Second, the three means do not need to be identifiable when the focus is only on  the difference between the two means $\mu_t - \mu_c$. In fact, by setting $\beta_t = \beta_c$, the difference between the two means, $\mu_t - \mu_c$, is identical to the difference between the two intercepts, $\tilde{\mu}_t - \tilde{\mu}_c$:
$$
\tilde{\mu}_t - \tilde{\mu}_c = (\mu_t - \beta_t \cdot \mu_0) - (\mu_c - \beta_c \cdot \mu_0) = \mu_t - \mu_c.
$$
This technique increases the power to detect a change between treatment and control, but it is not suited for inference on the percent change scale because the means are not separately identifiable. In contrast, working within a Bayesian framework makes the identification of the three means straightforward. This is discussed further in Section \ref{sec:met}.

\section{Methodology}\label{sec:met}

\subsection{The Pre-Post Model}\label{sub:mod}
 The approach described in this section combines the percent change interpretability of the methods described in Section~\ref{sec:classic} with the increase in statistical power of the control-variate variance reduction technique described in Section~\ref{sub:var}.

The model, which will be called Pre-Post in the remainder of the paper, has two components. The first component describes the observations in the pre-period.
\begin{equation}\label{eq:model}
X_{i, j}  \sim N(\mu_0, \sigma_0^2),
\end{equation}
where $i = 1, \ldots, n_j$ and $j=t,c$.  The mean and the variance are identical for the treatment and the control since no treatment has occurred yet. 

The second component describes the observations in the post-period given the pre-period
\begin{equation}\label{eq:model2}
Y_{i,j} | X_{i,j} = x_{i,j}, \mu_0  \sim N \big( \mu_j + \beta_j \cdot (x_{i,j} - \mu_0), \tau_j^2 \big), 
\end{equation}
where $i = 1, \ldots, n_j$ and $j=t,c$.
Given $\bm{x}_j$ and $\mu_0$, $Y_{i,j}$ follows a simple linear regression with slope $\beta_j$, independent variable $x_{i,j} - \mu_0$ and residual error $\epsilon_{i,j} \sim N(0, \tau_j^2)$, for each of the two groups $j=t,c$.
 
When using a Bayesian model, one must specify a prior for the unknown parameters. I consider the following improper priors
$$
\pi(\mu_0, \sigma_0^2, \mu_t, \beta_t, \tau_t^2, \mu_c, \beta_c, \tau_c^2)  \propto 1 / \sigma_0^2 \cdot 1 / \tau_t^2 \cdot 1 / \tau_c^2.
$$

Priors \eqref{eq:pri} are also called matching priors because the resulting CIs for $\mu_c$ and $\mu_t$ \emph{match}  the frequentist confidence intervals for $\mu_c$ and $\mu_t$, respectively, for all significance levels $\alpha \in (0, 1)$. In addition, these priors do not require any prior elicitation from the experimenter, and the resulting posterior distribution is invariant under affine transformations of the data. These properties are very convenient when an estimation method is integrated in an automated experiment framework that is used by many experimenters and on a large number of experiments and metrics such as we have at Google. An alternative, not explored in this work, is to use the class of objective priors proposed by \cite{ghosh2003objective} for the estimation of the ratio of regression parameters. 
If the experimenter instead has prior knowledge on the parameters of the model,  the objective prior \eqref{eq:pri} can be easily replaced with an informative prior.

The posterior distribution of $(\mu_t, \mu_c)$ can be easily computed using a Gibbs sampler. However, a Gibbs sampler can be computationally inefficient, particularly when one is interested in computing the posterior for a large number of metrics and experiments, as is the case in practice. A common approach used to overcome the computational limitation of the Gibbs sampler is variational inference. However, variational inference tends to underestimate variance, and in the context of online experimentation it is important to have an accurate estimate of the false positive rate, which will depend on an accurate estimate of the variance.  
To overcome the limitations of both the Gibbs Sampler and variational inference, I developed a fast and deterministic algorithm which discretely approximates the posterior of $\mu_t, \mu_c$ on a grid. The algorithm is described in Section~\ref{sub:alg}, and a comparison of the proposed algorithm to a Gibbs sampler is provided in  Section~\ref{sec:gibbs}. 
 
\subsection{Algorithm}\label{sub:alg}
In this section I describe an algorithm to approximate the posterior distribution of the Pre-Post model. The joint posterior distribution of $(\mu_t, \mu_c)$ can be written as follows
$$
\pi(\mu_t, \mu_c | \bm{x}_t, \bm{x}_c, \bm{y}_t, \bm{y}_c )  = \int 
\bigg[ \prod_{j=t,c} \pi(\mu_j | \bm{x}_j, \bm{y}_j, \mu_0) \bigg] 
\pi(\mu_0 | \bm{x}_t, \bm{x}_c, \bm{y}_t, \bm{y}_c  ) d \mu_0, 
$$
where $\bm{x}_j = (x_{1,j}, \ldots, x_{n_j, j})'$  and $\bm{y}_j = (y_{1,j}, \ldots, y_{n_j, j})'$ for $j = t,c$.

There is no closed form for the exact posterior of $\mu_0$, $\pi(\mu_0 | \bm{x}_t, \bm{x}_c, \bm{y}_t, \bm{y}_c )$. Instead of simulating the exact posterior of $\mu_0$ through a Gibbs Sampler, one can achieve a closed form approximation by a pseudo-posterior which is computed only with respect to the pre-period data, i.e., $\pi(\mu_0 | \bm{x}_t, \bm{x}_c)$. Unlike the exact posterior, the distribution of this pseudo-posterior can be derived analytically.

This pseudo-posterior will be close to the exact posterior because most of the information about $\mu_0$ is contained in the pre-period.
In fact, the pre-period mean $\mu_0$ cannot be identified from the likelihood of the post-period  \eqref{eq:model2} alone. This is due to the fact that in the likelihood \eqref{eq:model2} there are three parameters $\mu_j, \beta_j$ and $\mu_0$ to describe a two dimensional space.  When adding the likelihood of the pre-period \eqref{eq:model}, the pre-period mean $\mu_0$ becomes identifiable. 

The information about $\mu_0$ in the post-period is stronger when the correlations between the pre-period and the post-period are large. This can be seen by writing the combined likelihood of $\mu_0$ from the  the pre-period \eqref{eq:model} and the post-period \eqref{eq:model2},
\begin{align}\label{eq:likelihood}
\begin{split}
L(\mu_0   & | \bm{y}_t, \bm{y}_c, \bm{x}_t, \bm{x}_c, \mu_t, \mu_c, \rho_t, \rho_c, \ldots)  \\
& \propto \big[  \prod_{i,j} N(x_{i,j} | \mu_0, \sigma_0^2) \big] \cdot  \prod_j \big[ \prod_{i} N(y_{i,j} | \mu_0 + \rho_j \cdot \dfrac{\sigma_j}{\sigma_0}(x_{i,j} - \mu_j), (1 - \rho_j^2) \cdot \sigma_j^2 ) \big] \\
& \propto  N\big( \mu_0|  \bar{x}, \dfrac{\sigma_0^2}{ n_c + n_t} \big) \cdot \prod_j N  \big(\mu_0| \bar{x}_j + \rho_j \cdot \dfrac{\sigma_0}{\sigma_j} (\mu_j - \bar{y}_j ), \dfrac{ \sigma_0^2 }{ n_j} \cdot \dfrac{1 - \rho_j^2}{ \rho_j^2} \big).
\end{split}
\end{align}

In the likelihood \eqref{eq:likelihood} the key parameters are the correlations $\rho_t$ and $\rho_c$. The two correlations control the shift in the means as well as how concentrated  the  likelihood of the post-period is with respect to the likelihood of the  pre-period. When $\rho_t = \rho_c \simeq 0$, for example, the shifts are minimal, and the post-period likelihood is much more dispersed than the pre-period likelihood. This implies that the information about $\mu_0$ is primarily concentrated in the pre-period likelihood.
The post-period means $\mu_j$ and variances $\sigma_j^2$ appear in the likelihood of the post-period, but they only play a role of centering $\bar{y}_j$ to zero and scaling it to the standard deviation of the pre-period. In fact, $\sigma_0 / \sigma_j \cdot (\mu_j - \bar{y}_j ) \sim N(0, \sigma_0^2 / n_j)$,  where $j=c,t$.

Since the pre-period likelihood \eqref{eq:model}
is the dominant anchor in the posterior of $\mu_0$, I posit that little information is lost by approximating it as 
$\pi(\mu_0 | \bm{x}_t, \bm{x}_c, \bm{y}_t, \bm{y}_c ) \simeq \pi(\mu_0 | \bm{x}_t, \bm{x}_c)$. This is validated in Section~\ref{sec:gs_vs_da}  with an analysis comparing the exact posterior
$\pi(\mu_0 | \bm{x}_t, \bm{x}_c, \bm{y}_t, \bm{y}_c )$ and pseudo-posterior $\pi(\mu_0 | \bm{x}_t, \bm{x}_c)$, as well as an analysis on the downstream impact on the inference on the percent change $100 \cdot \mu_t / \mu_c - 100$. Specifically, since the key parameters in \eqref{eq:likelihood} are the correlations $\rho_t$ and $\rho_c$, the analysis focuses on the impact of the correlations on the accuracy of the inference.

Given the pseudo-posterior for $\mu_0$, I now outline an approach to computing the posterior $\pi(\mu_t, \mu_c | \bm{x}_t, \bm{x}_c, \bm{y}_t, \bm{y}_c ) $.
A continuous univariate distribution can be discretely approximated by computing its quantiles. For example, the $d$th element of the discrete approximation will correspond to the quantile $(2 \cdot d - 1) / (2 \cdot D)$, and each element will have equal probability $1/D$. By using such discretization, one can deterministically approximate the joint posterior distribution of $\mu_0, \mu_c$ and $\mu_t$. 

Altogether, this gives us,
\begin{align}\label{eq:disc}
\begin{split}
\pi(\mu_t, \mu_c | \bm{x}_t, \bm{x}_c, \bm{y}_t, \bm{y}_c ) &  \simeq 
 \int \big[ \prod_{j=t,c} \pi(\mu_j | \bm{x}_j, \bm{y}_j, \mu_0) \big] \pi(\mu_0 | \bm{x}_t, \bm{x}_c) d \mu_0 \\
& \simeq \dfrac{1}{D} \sum_{d_0=1}^{D} \prod_{j=t,c} \pi(\mu_j | \bm{x}_j, \bm{y}_j, \mu_0^{(d_0)}) \\
& \simeq \dfrac{1}{D^3} \sum_{d_0=1}^{D} \prod_{j=t,c} \sum_{d_j=1} ^ D \delta( \mu_j^{(d_j)} | \bm{x}_j, \bm{y}_j, \mu_0^{(d_0)} ),
\end{split}
\end{align}
where $\delta(\cdot)$ represents the Dirac delta function, $ \mu_0^{(1)}, \ldots, \mu_0^{(D)}$  represent the discretization of the pseudo-posterior distribution of 
$\mu_0$, and 
$$ (\mu_j^{(1)} | \bm{x}_j, \bm{y}_j, \mu_0^{(d_0)} ), \ldots, (\mu_j^{(D)} | \bm{x}_j, \bm{y}_j, \mu_0^{(d_0)} )$$
 indicate the discretization of the conditional distributions of  $\mu_j$ given $\mu_0 = \mu_0^{(d_0)}$  for $j = c,t$. 

Definitions for the posterior distributions of the pre-period mean $\pi(\mu_0 | \bm{x}_t, \bm{x}_c)$ and post-period means $\pi(\mu_j | \bm{x}_j, \bm{y}_j, \mu_0)$ for $j=t,c$ are the final steps needed to complete the derivation of $\pi(\mu_t, \mu_c | \bm{x}_t, \bm{x}_c, \bm{y}_t, \bm{y}_c ) $.

\paragraph{Pre-period mean estimation} Under the matching prior $\pi(\mu_0, \sigma_0^2) \propto 1 / \sigma_0^2 $, the distribution of the pseudo-posterior of $\mu_0$ is equal to
 $$
 T_{n_c + n_t - 1}(\bar{y}_0, s_0),
 $$
 where $\bar{y}_0$ represents the sample mean of the two groups combined in the pre-period, and $s_0$ indicates the sample standard deviation of the two groups combined in the pre-period, and $T_x(y, z)$ indicates a non-standardized Student's t-distribution with $x$ degrees of freedom, location parameter $y$ and scale parameter $z$.
  
\paragraph{Experiment period means estimation}
As discussed earlier, given $\bm{x}_j$ and $\mu_0$, $Y_{i,j}$ follows a simple linear regression with independent variable $x_{i,j} - \mu_0$ and residual variance $\tau_j^2$, for each of the two groups $j=t,c$.

Recall that the priors for $(\mu_c, \beta_c, \tau_c^2)$ and $(\mu_t, \beta_t, \tau_t^2)$ are mutually independent. Thus,  
given $\mu_0$, the posterior distributions of $\mu_c$ and $\mu_t$ are also mutually independent. 
Specifically, given $\mu_0$, the posterior for $\mu_j$ is equal to
\begin{align*}
\mu_j  &| \bm{x}_j, \bm{y}_j, \mu_0  \sim T_{n_j - 2}(\hat{\mu}_j , \hat{\tau}_j  \cdot z_j) \\
z_j & := \big( \sum_{i = 1}^{n} x_{i,j} ^ 2 ) / \big(n_j \cdot (n_j - 1) \cdot s_{0,j} ^ 2 \big),
\end{align*}
where  $\hat{\mu}_j $ is the ordinary least squares estimate of $\mu_j$,  $\hat{\tau}_j$ is the ordinary least squares estimate of the standard deviation of the residuals, and  $s_{0,j}^2$ is the sample variance in the pre-period for group $j$, where $j = c,t$. Note that $\hat{\mu}_j = \hat{\mu}_j(\bm{x}_j, \bm{y}_j, \mu_0)$  and $\hat{\tau}_j = \hat{\tau}_j(\bm{x}_j, \bm{y}_j, \mu_0)$, since they are estimates of a simple linear regression with dependent variable $y_{i,j}$ and independent variable equal to $x_{i,j} - \mu_0$.
 
As described in \eqref{eq:disc} , the distributions of the means of the control group and the treatment group are discretized based on their quantiles. Given the pre-period mean, the two distributions are conditionally independent, and their joint distributions can be approximated by the cross product of the two discrete approximations.

\begin{figure}[ht]
  \centerline{\includegraphics[width=\textwidth]{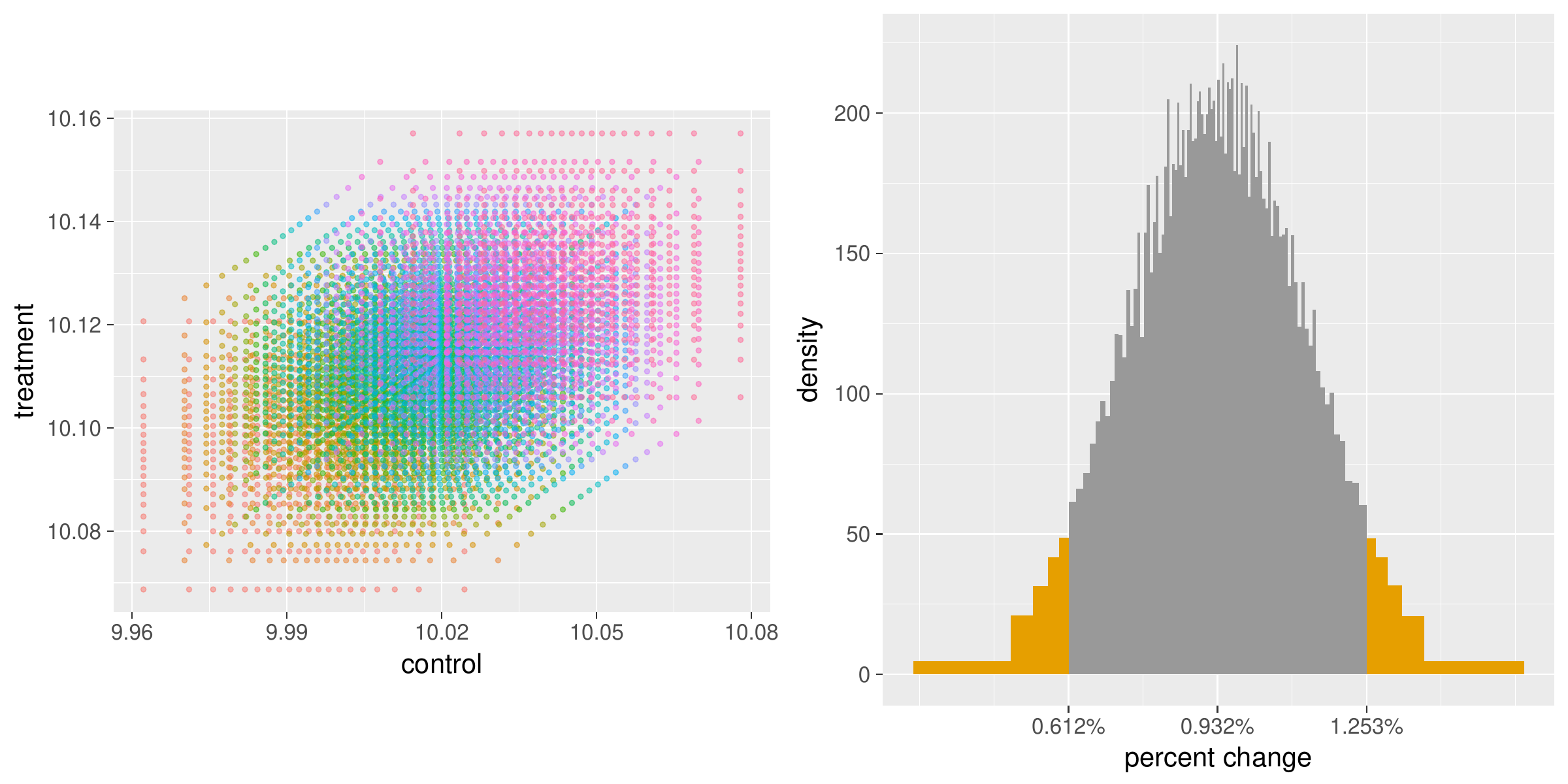}}
  \caption{In the scatterplot on the left each dot indicates a pair $(\mu_t^{(d)}, \mu_c^{(d)})$} for $d = 1, \ldots, D ^ 3 = 20 ^ 3$. The dots are colored based on the 20 values $\mu_0^{(1)}, \ldots, \mu_0^{(20)}$. The histogram on the right indicates the approximation to the posterior distribution for the percent change between $\mu_t$ and $\mu_c$ based on the  $20 ^ 3$ pairs $(\mu_t^{(d)}, \mu_c^{(d)})$. Each of the 100 bins has equal probability. The 90 central bins are colored in grey to visually identify, as an example, the 90\% CI.  
  \label{fig:gri}
\end{figure}

Figure~\ref{fig:gri} shows how the discrete approximation of $\pi(\mu_t, \mu_c | \bm{x}_t, \bm{x}_c, \bm{y}_t, \bm{y}_c ) $ is constructed through a toy example based on a simulated dataset. The scatterplot on the left shows the pairs $(\mu_t^{(d)}, \mu_c^{(d)})$ for $d=1,\ldots, D ^ 3 = 20 ^ 3$, and the histogram on the right shows the percent change $100 \cdot \mu_t^{(d)} / \mu_c^{(d)} - 100$, and the estimate of the 90\% CI based on the sample quantiles. 

\section{Empirical Results}\label{sec:emp}

This section is organized in three sub-sections. In the \ref{sec:nodes} I compare the deterministic algorithm (DA) presented in Section \ref{sub:alg} and the Gibbs sampler (GS). In \ref{sec:gs_vs_da} I quantify the benefits of Pre-Post with respect to an equivalent model that does not include the pre-period using a large number of YouTube A/A experiments. In \ref{sec:real} I compare Pre-Post  with existing methods on simulated data. 

\subsection{Comparison of the Deterministic Algorithm and the Gibbs Sampler}\label{sec:gibbs}

Three examples are presented in this section.
The goal of these three examples is to show that the DA is comparable to the GS in term of inference, but that it is substantially more efficient computationally.

In the first example I study the robustness of the estimates as a function of the number of nodes $D$ for the DA and the number of iterations for the GS. 
In the second example I compare the parameter estimates of the DA and the GS.
In the third example I compare the computing times of the DA and the GS.

\subsubsection{Number of nodes and number of iterations}\label{sec:nodes}

Both the DA and the GS algorithms have some parameters to be chosen to obtain stable and reliable inference. The parameter of the DA is the number of nodes, and the parameters of the GS are the burn-in and the number of iterations. The goal of this analysis is to identify  good parameter values for the two algorithms.

\begin{figure}[ht]
  \centerline{\includegraphics[width=\textwidth]{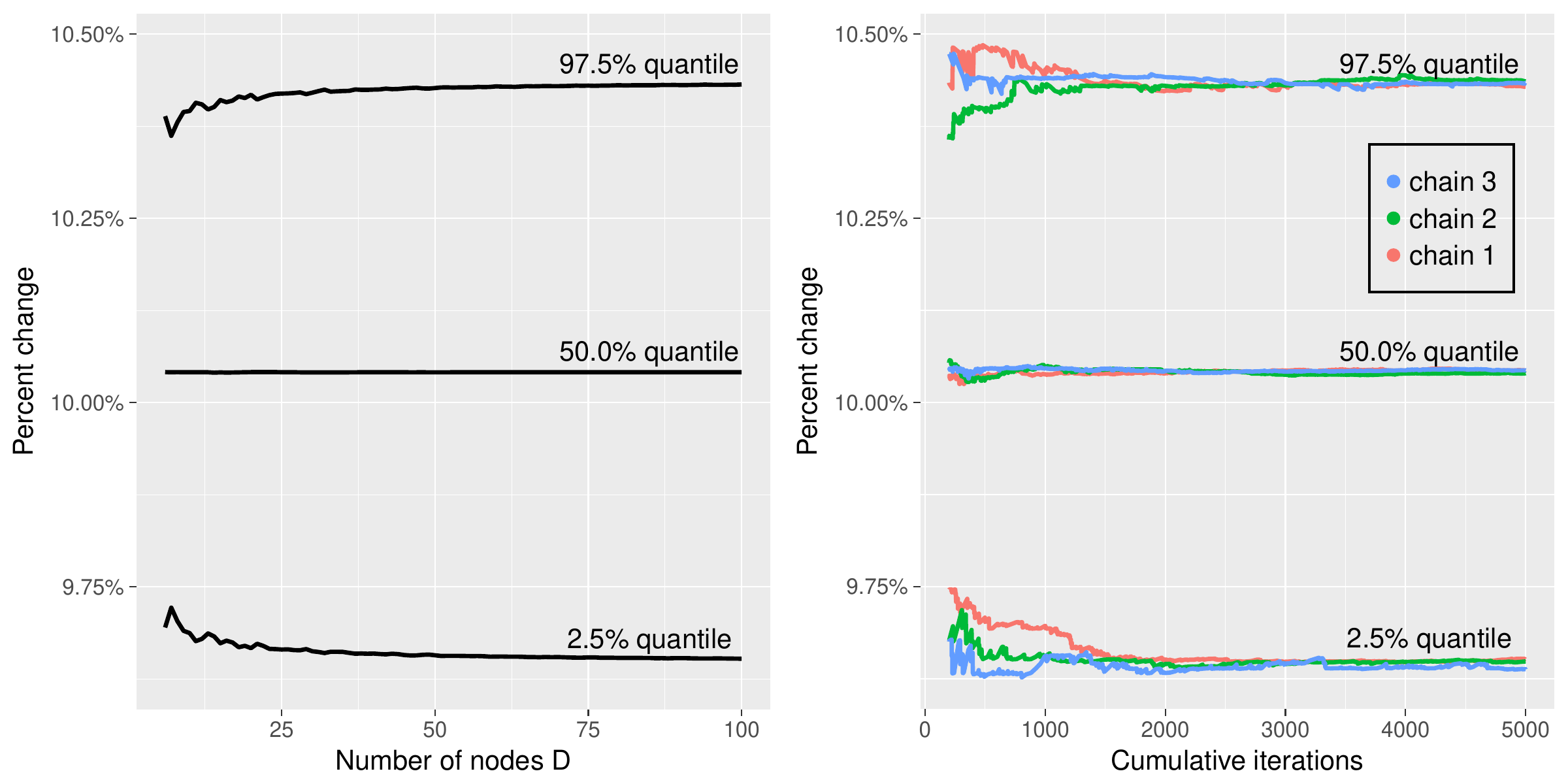}}
  \caption{For a simulated dataset, the plot on the left shows the estimates of 97.5\%, the 50.0\% and the 2.5\% quantiles of the percent change in function of the number of nodes $D$. The plot on the right shows the same estimates based on three Markov chains, where each color indicates a different Markov chain.}
  \label{fig:mcmc}
\end{figure}

For a single simulated dataset, the plot on the left of Figure~\ref{fig:mcmc} shows the estimates of the 97.5\%, the 50.0\% and the 2.5\% quantiles of the percent change \eqref{eq:per} as a function of the number of nodes $D$. The estimate of the 50\% quantile is immediately stable, while the estimates of the other two quantiles are stable starting from $D \simeq 50$. 

The plot on the right shows the estimates of the same quantiles based on three independent Markov chains, where different colors are used to identify the different Markov chains. The Markov chains reached a stationary state within the first 100 iterations. For this reason the chains are computed using a GS with a burn-in of 100 iterations. As one can see from the plot, once the Markov chains have accumulated $\simeq 2000$ iterations, the estimates of the quantiles are stable. 

Based on the results of this analysis I fixed  the number of iterations of the GS to 2000 and the number of nodes of the DA to 50 for all of the simulations of the empirical results section.

\subsubsection{Estimation}\label{sec:gs_vs_da}
As shown in \eqref{eq:likelihood} the key parameters impacting the the accuracy of the pseudo-posterior in the DA algorithm are the correlations $\rho_c$ and $\rho_t$. To understand the accuracy of the pseudo-posterior in this example I compare the parameter estimates of the DA and the GS as a function of the correlations between pre-period and post-period. 
 Specifically, I compare the posterior mean and the posterior standard deviation of $\mu_0$ and  the posterior mean and the posterior standard deviation of  $100 \cdot \mu_t / \mu_c - 100$ between the DA and the GS over a class of simulated datasets indexed by $\rho_t = \rho_c = \rho$.  The datasets are simulated from the following model

\begin{eqnarray}\label{eq:bivariate_example}
\begin{pmatrix}X_{i,j}\\
Y_{i,j}
\end{pmatrix} & \sim & N\left[\left(\begin{array}{c}
100\\
100 + 10  I_{(j = t)}
\end{array}\right),\left(\begin{array}{cc}
1 & \rho \\
\rho & 1 
\end{array}\right)\right],
\end{eqnarray}

where $\rho =  0, 0.4, 0.8$, $i = 1, \ldots, 50$ and $j = c,t$. For each correlation value I simulated 50 datasets.
Similar results are obtained when $\rho_t \neq \rho_c$ or when $\sigma_0^2 \neq \sigma_t^2 \neq \sigma_c^2$, and so they are not presented here.

\begin{figure}[ht]
  \centerline{\includegraphics[width=\textwidth]{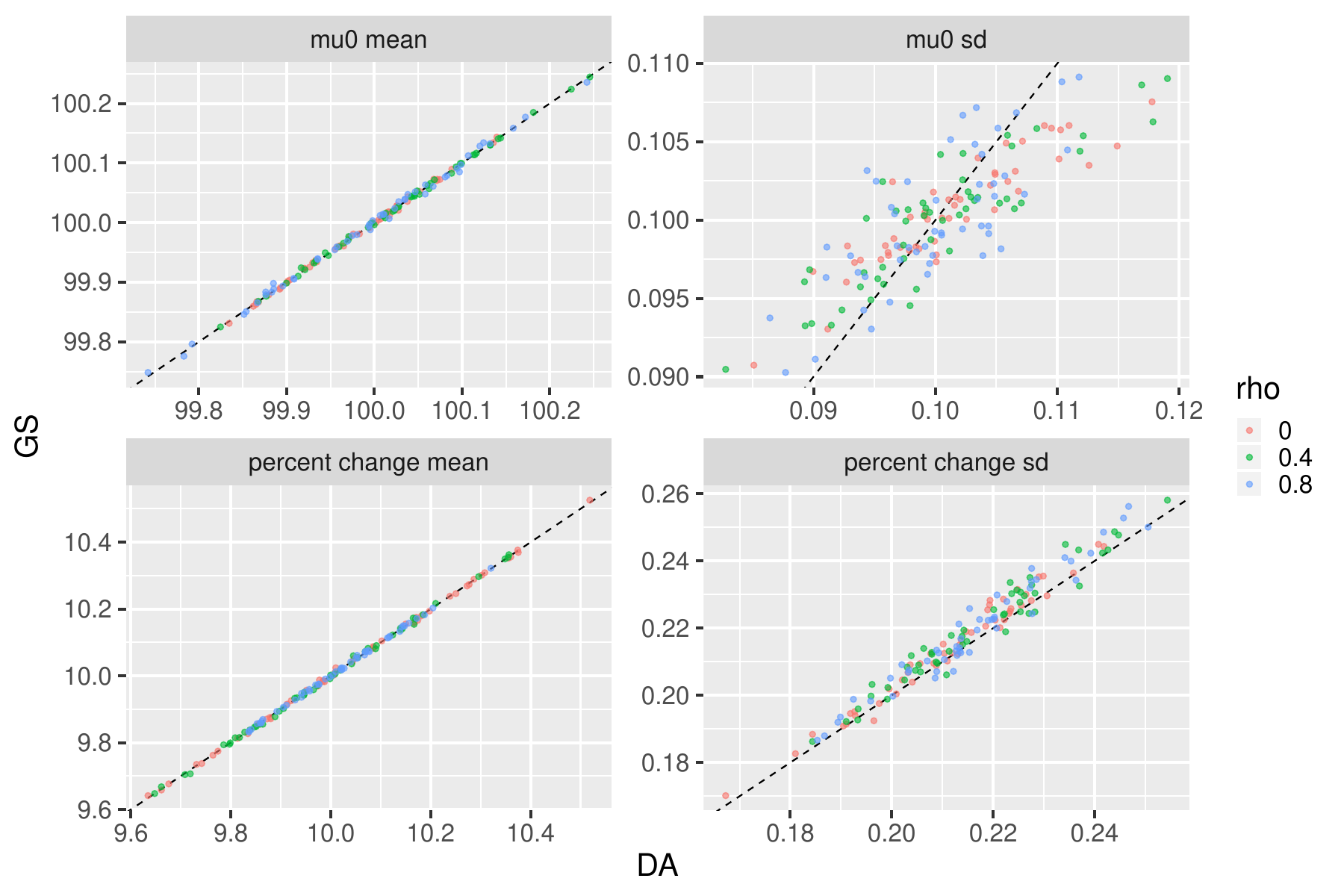}}
  \caption{Comparison between the inference based on the GS on y-axis and the DA on the x-axis. Each dot represents a simulated datasets. The color of the dot indicates the correlation level $\rho$. Red indicates $\rho = 0$, green indicates $\rho = 0.4$, and blue indicates $\rho = 0.8$. The plot of the top left shows the posterior mean of $\mu_0$, the plot on the top right shows the posterior mean of $\mu_0$. The plot on the bottom left shows the posterior mean of  the percent change $100 \cdot \mu_t / \mu_c - 100$. 
  The plot on the bottom right shows the posterior standard deviation of  the percent change $100 \cdot \mu_t / \mu_c - 100$ scaled by $\sqrt{1 - \rho^2}$.}
  \label{fig:gs_da}
\end{figure}

The scatterplots in Figure~\ref{fig:gs_da} show the comparison  between the DA and the GS for the posterior mean and the posterior standard deviation of $\mu_0$, and for the posterior mean and the posterior standard deviation of  the percent change $100 \cdot \mu_t / \mu_c - 100$. The posterior standard deviation of  the percent change has been scaled by $\sqrt{1- \rho^2}$ to make the results comparable across different correlation levels. 

The scatterplots  show that the posterior mean of $\mu_0$ is practically identical between the two algorithms. The standard deviation of $\mu_0$ has higher variability when using the DA algorithm, but both algorithms are centered around the same value. This means that there is no systematic bias in the estimation of the posterior standard deviation of $\mu_0$ between the DA and the GS.

The posterior mean of the percent change $100 \cdot \mu_t / \mu_c - 100$ is practically identical between the two algorithms.  The standard deviation of the percent change $100 \cdot \mu_t / \mu_c - 100$ is slightly larger for the GS algorithm, but the difference is minimal. 

Based on the scatterplots  I can conclude that the correlation level does not seem to impact the accuracy of the inference.  

\subsubsection{Computing Time}

In this example I compare the  computing times of the DA and the GS over 100 datasets. 
The datasets are simulated from model \eqref{eq:bivariate_example}, where $\rho = \text{cor}(X_{i,j}, Y_{i,j}) = 0.8$, $i = 1, \ldots, 20$ and $j = c,t$.

The average computing time for the DA is 0.16 seconds, while the average computing time for the GS is 36.56 seconds. This means that the DA results in a 200x reduction in computing time with respect to the GS. The code was written in R for both approaches, and it was tested on an Intel Xeon CPU E5-1650 v3 at 3.50GHz. 

\subsection{Real Data Examples}\label{sec:real}

The goal of this section is to quantify the benefits of Pre-Post in real YouTube experiments.

To better understand the benefits of incorporating a pre-period metric, I compare Pre-Post to a baseline model which does not include the pre-period metric. 
The model, which will be called the Post method in the remainder of this work, is based on the likelihood \eqref{eq:post} and the following prior
the 
 \begin{equation}\label{eq:pri}
\pi(\mu_c, \sigma_c^2, \mu_t, \sigma_t^2) \propto 1 / \sigma_c^2   \cdot 1 / \sigma_t^2.
 \end{equation}

In the first example I study the robustness of the methodology as a function of the level of misalignment between the control and treatment group in the pre-period. 
In the second example I study the width of the CIs as a function of the traffic size and the number of days of the experiment. Both examples are based on the DAUs metric. Similar results are obtained for other metrics and are for this reason omitted.

\subsubsection{Robustness to Pre-Period Misalignment}

The experiment framework is designed not to have a systematic pre-period bias between the groups of an experiment. However, engineers often run A/A experiments prior to the start of their experiments out of the concern that the two groups are not perfectly balanced. 
The original goal of this analysis was to convince engineers at YouTube that Pre-Post is robust to the natural variability in the level of pre-period alignment between the two groups. 

This example is based on 100,000 A/A YouTube experiments. A/A experiments are used for two reasons. First, to assess the performance of the methods a very large number of experiments are needed, and it is easy to generate a large number of A/A experiments. Second, when using A/A experiments the true percent change is known, making it possible to assess both the coverage of CIs and the mean squared error (MSE) of the point estimates. 

I compute the permutation \pvalue\ for the mean difference in the pre-period, and then I bucket all the experiments in 100 groups based on their pre-period \pvalue. The first group contains all experiments where the pre-period \pvalue\ $\in [0, 0.01)$, the second groups contains all experiments where the pre-period \pvalue\ $\in [0.01, 0.02)$, etc. 
In the pre-period there is no treatment applied to the treatment group, and so any difference in the distribution of the two groups is due to chance. Thus, the pre-period \pvalue\ has a uniform distribution, resulting in approximately 1,000 experiments per pre-period \pvalue\ bucket.

\begin{figure}[ht]
\centerline{\includegraphics[width=\textwidth]{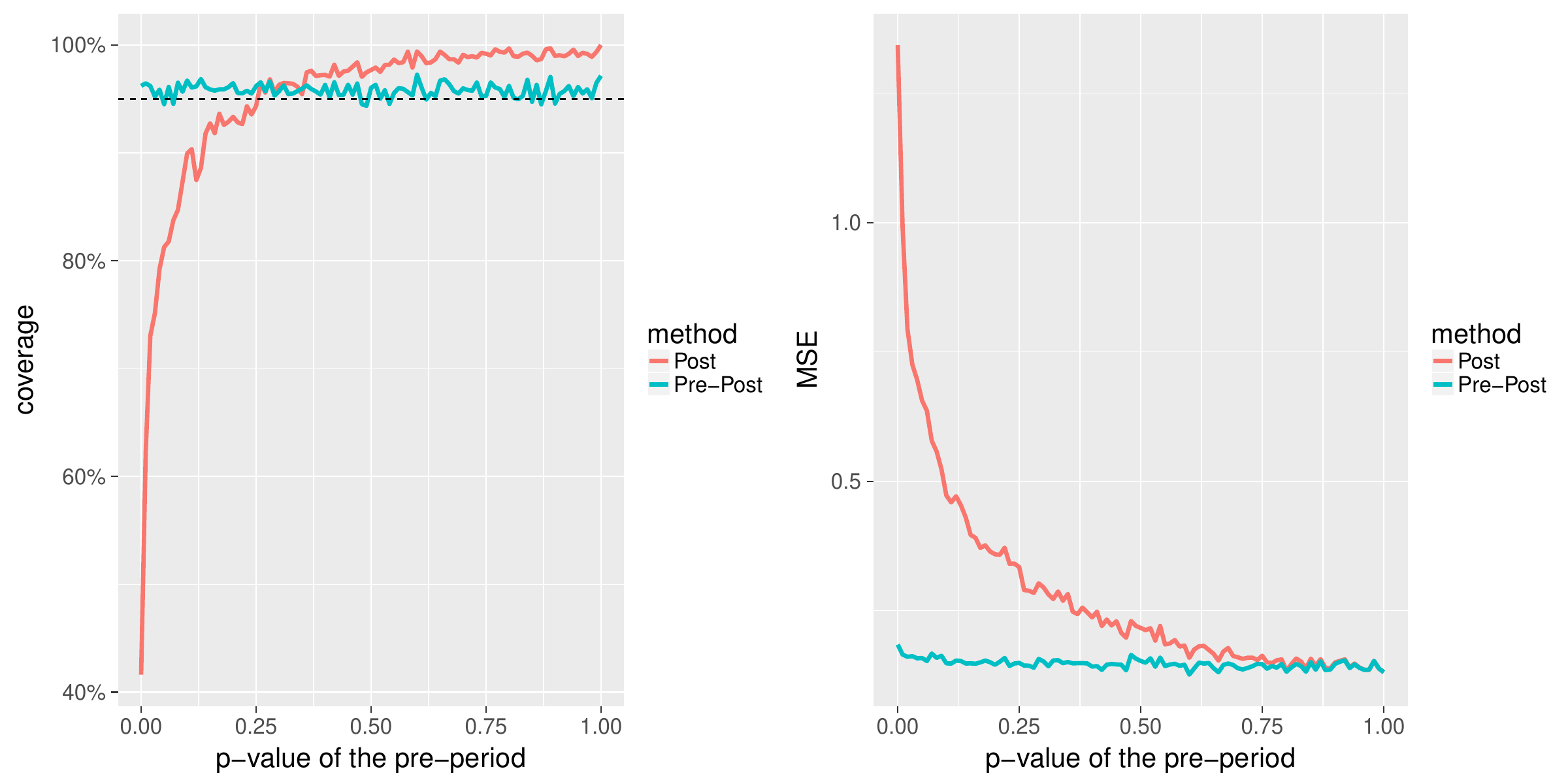}}
 \caption{On the left, coverage of the CIs in the post-period as a function of the permutation \pvalue\ for the pre-period.  On the right, MSE for the post-period as a function of the permutation \pvalue\ for the pre-period. The red line indicates Post, and the blue line indicates Pre-Post. The dashed black line in the left plot indicates 95\%, the nominal coverage of the CIs.}
\label{fig:mis}
\end{figure}

For each pre-period \pvalue\ bucket Figure~\ref{fig:mis} shows the empirical coverage
 of the  95\% CI and the MSE in the post period. Pre-Post outperforms Post in both coverage robustness and MSE. Both Pre-Post and Post have the right coverage on average. However, the coverage of Pre-Post is uniform with respect to the pre-period \pvalue, while the coverage of Post is strongly dependent on the pre-period permutation \pvalue. In particular, when the pre-period \pvalue\ is smaller than 0.1, i.e., the two groups are not well aligned in the pre-period, Post has substantially lower coverage than the nominal coverage. Similarly, the MSE of Pre-Post is uniform, while the MSE of Post is very high when the pre-period \pvalue\ is small and comparable to Pre-Post when the \pvalue\ is large.

\subsubsection{Width of the Credible Intervals}

There is a common misconception that running an online experiment for twice the time is equivalent to running and experiment on twice the traffic. Similar misconceptions have been noticed by data scientists at other tech companies \cite{xu2018sqr}. Since finding space for a large experiment is often difficult, experimenters often run their experiments longer to compensate for the small amount of traffic available. 

The goal of this example is to illustrate the relationship between the size of an experiment at YouTube, its length and the method used for inference. In particular, the example shows how running experiments twice as long is not as valuable as running experiments on twice the traffic. More importantly, the example shows that a variance reduction technique like Pre-Post can be even more valuable than doubling the size of the experiment.

This example is based on 40 identical YouTube experiments, where 20 experiments have twice the traffic ($2p$) of the other 20 experiments ($p$). The pre-period corresponds to the 7 days prior to the start of the experiment. The metric considered is DAUs.

\begin{figure}[ht]
  \centerline{\includegraphics[width=\textwidth]{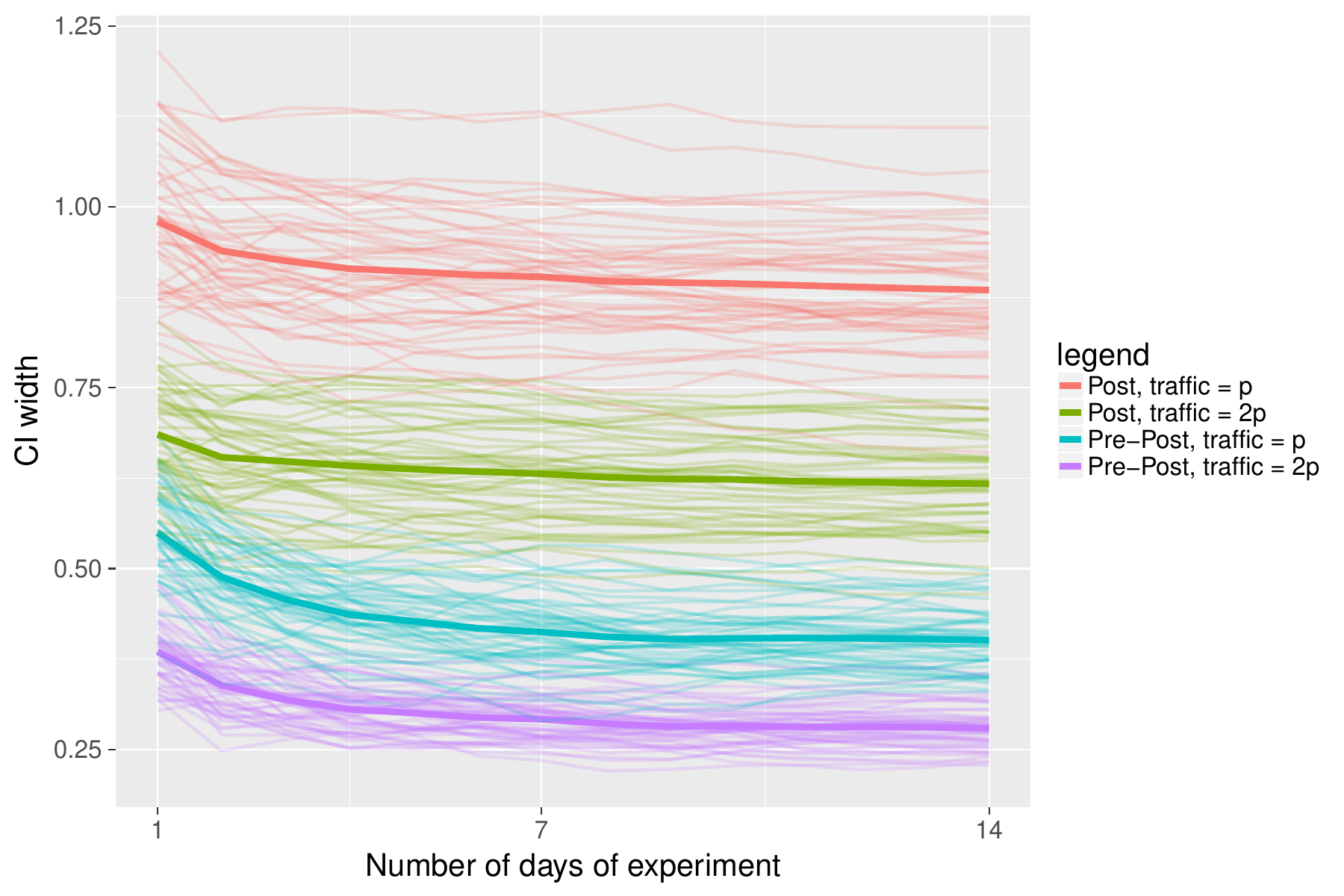}}
  \caption{Width of CIs as a function of the number of days of the experiment. Red corresponds to Post based on traffic proportion $p$, green corresponds to Post based on traffic proportion $2p$, blue corresponds to Pre-Post based on traffic proportion $p$, and purple corresponds to Pre-Post based on traffic proportion $2p$. The thin lines indicate individual experiments, and the thick lines indicate the average across experiments.}
  \label{fig:wid}
\end{figure}

In Figure~\ref{fig:wid}, each thin line represents an individual experiment, and the thick line represents the average across experiments. Each color represents a different method and traffic size combination. Specifically, red corresponds to Post based on traffic proportion $p$, green to Post based on twice the traffic ($2p$), blue to Pre-Post based on traffic proportion $p$, and purple to Pre-Post based on twice the traffic ($2p$). 

Correcting for the pre-period results in substantially tighter CIs. In this example the CIs for Pre-Post with traffic proportion $p$ are tighter than the CI of Post with traffic proportion $2p$. In addition, even if one takes into account the week spent to run the pre-period, the CIs of Pre-Post are still substantially tighter than the CIs of Post. Specifically, the CIs of Pre-Post after 1 day of the experiment are tighter than the CIs of Post based on 8 or more days of the experiment. 

The rate at which the width of the CIs decays is slower than the inverse of the square root of the number of days. This is due to the fact that metrics are generally not independent across days, but instead are positively correlated over time. This implies that unless the treatment effect increases over time, the gains of running a longer experiments are smaller than those of running a larger experiment. Given the fact that a large number of experiments are often running in parallel (\cite{tang2010overlapping}), it is not possible to have a large sample size for each of them. Thus, maximizing the sensitivity of each individual experiment through methods like Pre-Post is crucial.

\subsection{Simulation Study}\label{sec:study}

In this simulation study Pre-Post and Post are compared to the following classical methods: Taylor's, Fieller's and the Index method.
Two generative models are considered. In the first model data are generated from a Bernoulli model to mimic a metric like DAUs.

\begin{align*}
\tilde{X}_{u,j} | p_{u,j} & \sim \text{Bernoulli}(p_{u,j}) \\
\tilde{Y}_{u,c} | p_{u,c} & \sim \text{Bernoulli}(0.9 \cdot p_{u,c}) \\
\tilde{Y}_{u,t} | p_{u,t} & \sim \text{Bernoulli}(0.9 \cdot (1 + \tau) \cdot p_{u,t}) \\
p_{u, j} & \sim \text{Beta}(0.2, 0.3),
\end{align*}
where $u = 1, \ldots, 10^5$ indicates the user, $j=c,t$ represents the condition, and $\tau$ represents the treatment effect. The latent variable $p_{u,j}$ is associated to the propensity of user $u$ to use the product on any given day.  The Beta distribution over the latent variables is bimodal to mimic two distinct ``types" of users. Recurring users, where $p_{u,j} \simeq 1$, are likely to use the product both in the pre-period and the post-period. Occasional users, where $p_{u,j} \simeq 0$, are likely not to use the product at all or use it only in the pre-period or the post-period. The parameter $\tau$ represents the increase in the propensity for a user in the treatment group to use the product in the post-period.

In the second model data are generated from an Exponential model to mimic a time metric, like time on site or latency.

\begin{align*}
\tilde{X}_{u,j} | \lambda_{u,j} & \sim \text{Exponential}(\lambda_{u,j}) \\
\tilde{Y}_{u,c} | \lambda_{u,c} & \sim \text{Exponential}(\lambda_{u,c}) \\
\tilde{Y}_{u,t} | \lambda_{u,t} & \sim \text{Exponential}(\lambda_{u,t} / (1 + \tau)) \\
1 / \lambda_{u, j} & \sim \text{Beta}(0.1, 0.9),
\end{align*}
where $u = 1, \ldots, 10^5$ indicates the user,  $j=c,t$ represents the condition, and $\tau$ represents the treatment effect. Similarly to the first model, the latent variable $\lambda_{u,j}$ is associated to the amount of time the user uses the product on any given day. The parameter $\tau$ represents the increase in the expected time for a user in the treatment group during the experiment with respect to a user in the control group. 

In both models data are subsequently aggregated into buckets $i=1, \ldots, I = 50$ by randomly assign users to buckets,

\begin{align*}
P(u \in \text{bucket}_i) & = 1 / I \\
X_{i,j} & = \sum_{u \in \text{bucket}_i} \tilde{X}_{u,j} \\
Y_{i,j} & = \sum_{u \in \text{bucket}_i} \tilde{Y}_{u,j}.
\end{align*}

For both models 11 treatment effects ranging from 0\% to 10\% are considered, and for each treatment effect level $10^4$ datasets are generated.

 Table~\ref{tab:coverage} shows the empirical coverage of the 95\% CIs for each model across the $11 \cdot 10 ^ 4$ data sets and under each generative model. All methods have very good nominal coverage. 

\begin{table}[ht]
\centering
\begin{tabular}{lrrrrrr}
  \hline
  &    \multicolumn{1}{c}{Bernoulli} & \multicolumn{1}{c}{Exponential} \\
  \hline
 Fieller & 0.953  & 0.953  \\ 
 Index & 0.952  & 0.953  \\ 
 Post & 0.948  & 0.948  \\ 
 Pre-Post & 0.952  & 0.951  \\ 
 Taylor & 0.948  & 0.948  \\ 
   \hline
\end{tabular}
\caption{Empirical coverage of the 95\% CIs  for each method and under each generative model across the $11 \cdot 10 ^ 4$ data sets.}\label{tab:coverage}
\end{table}

\begin{figure}[ht]
  \centerline{\includegraphics[width=\textwidth]{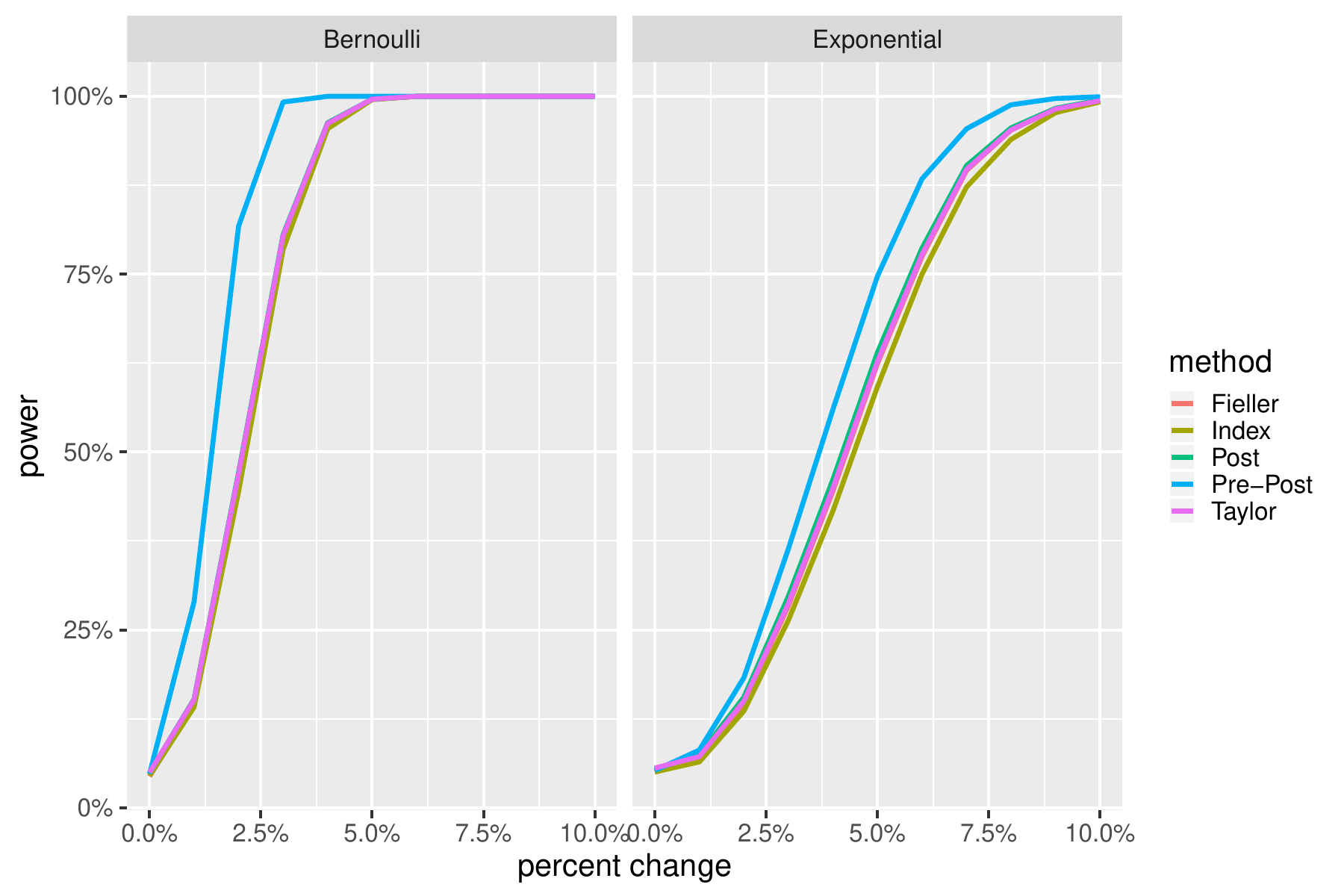}}
  \caption{Power as a function of the treatment effect on the percent change scale. The plot on the left shows the results for the Bernoulli generative model, and the plot on the right shows the results for the Exponential generative model. Red indicates the Fieller's method, brown indicates the Index method, green indicates the Post method, blue indicates the Pre-Post method, and purple indicates the Taylor's method.}
  \label{fig:sim_power}
\end{figure}

Figure~\ref{fig:sim_power} shows the power as a function of the treatment effect on the percent change scale for the two generative models. Each line in the plots represents a different method. For both generative models Pre-Post (blue line) is uniformly the most powerful approach, and all other methods have very similar power to one another.

\begin{figure}[ht]
  \centerline{\includegraphics[width=\textwidth]{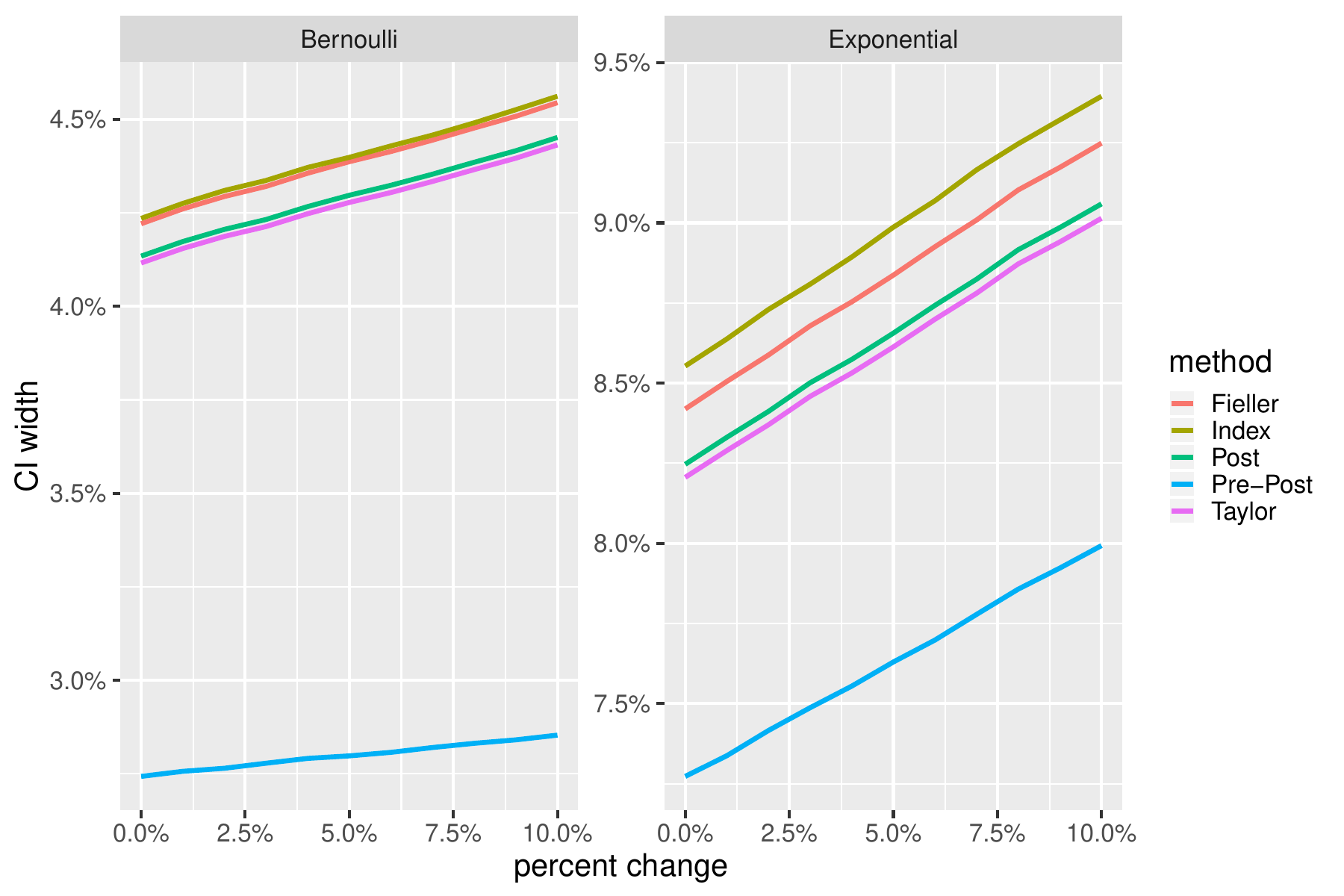}}
  \caption{CI width as a function of the treatment effect on the percent change scale. The plot on the left shows the results for the Bernoulli generative model, and the plot on the right shows the results for the Exponential generative model. Red indicates the Fieller's method, brown indicates the Index method, green indicates the Post method, blue indicates the Pre-Post method, and purple indicates the Taylor's method.}  
  \label{fig:sim_wid}
\end{figure}

Figure~\ref{fig:sim_wid} shows  the average width of the a CIs as a function of the treatment effect on the percent change scale for the two generative models. Each line in the plots represents a different method. For both generative models Pre-Post (blue line) produces the tightest CIs, while the Index method produces the widest CIs. The reduction in CI width of Pre-Post versus the other methods is larger under the Bernoulli model because the correlation between pre-period and post-period is higher in the Bernoulli model ($\sim0.8$) than in the Exponential model ($\sim0.5$). The width of the CIs grows approximately linearly in the increase in the percent change, consistently with the equation for the standard error of the Taylor's model \eqref{eq:taylor}.

\begin{figure}[ht]
  \centerline{\includegraphics[width=\textwidth]{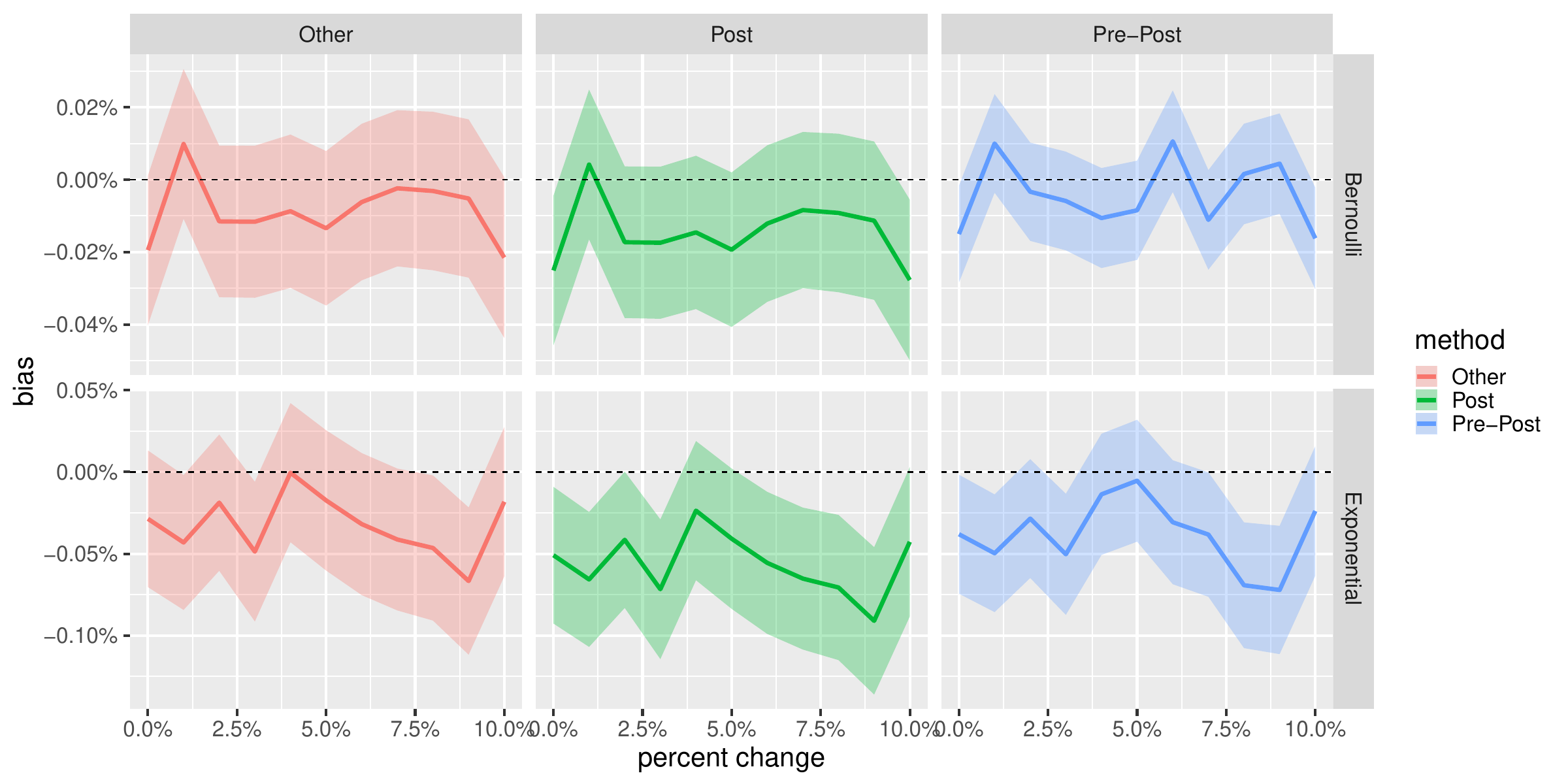}}
    \caption{Average bias and the associated 95\% CI as a function of the treatment effect on the percent change scale. The plots on the first row show the results for the Bernoulli generative model, and the plots on the second row show the results for the Exponential generative model. The first column (red) shows the results for the Fieller's method, the Index method, and the Taylor's method combined. For these methods the results are combined because they all share the same point estimate. The second column (green) shows the results for the Post method, and the third column (blue) shows the results for the Pre-Post method.}  
  \label{fig:sim_bias}
\end{figure}

 It is important to make sure that the gains in variance reduction do not result in a significant increase in bias. 
 One of the main criticisms of Bayesian models is that they do not result in unbiased estimates.
 I computed the average bias across the $10^4$ datasets and the associated 95\% CI of all the methods under each treatment effect level and each generative model. The results in Figure~\ref{fig:sim_bias} show a similar minimal downwards bias across all methods, i.e., the point estimate is smaller than the true percent change. There is no bias under the Bernoulli generative model and the bias is negligible with respect to the CI width for all methods under the Exponential generative model and uniformly over the size of the treatment effect.

\section{Conclusion}\label{sec:con}

Pre-Post can substantially increase the sensitivity of large scale online experiments relative to existing approaches. 
The width of CIs associated with YouTube experiments, for example, can be reduced by up to 50\%. 
This creates the opportunity for faster experimental cycles through shorter experiments, or exposing a smaller fraction of users to the experiment, while maintaining the same statistical power. Similarly, one can substantially increase the power to detect small effects, while maintaining the same length of the experiment and the same traffic size. Detecting small effects is important for web-facing products where the sum of several small improvements on a relative scale may have an overall large impact on an absolute scale. 

The main limitation of Pre-Post is the necessity to run a pre-period for a week before the start of the experiment. To overcome this limitation, YouTube developed an experiment framework to retrospectively compute pre-periods. Under this framework, for all users exposed to the treatment or the control group in the post-period, the pre-period metric $X$ is automatically and retrospectively computed. As a result, the experimenter does not need to set up a pre-period and can immediately start the experiment, resulting in simpler and faster experimental cycles.

Pre-period metrics can also be used to constantly monitor the health of the experiment system. Systematic deviations in the pre-period metrics between the treatment and the control can be an indication, for example, of issues in the traffic diversion or the logging.

There are many ways this methodology can be further extended. Often experimenters have run similar experiments in the past, and so they have prior knowledge about the range of possible values for the percent change between the treatment mean and the control mean. Parametrizing the model as a function of the control mean and the relative change of the treatment mean versus the control mean would make it easier to incorporate this prior knowledge. This parametrization also opens the door for heterogenous treatment effect modeling, which has been an area of very active research in the last few years \cite{athey2016recursive}, \cite{deng2016concise},\cite{wager2018estimation} and \cite{xie2018false}. Since the percent change is scale-free, it is reasonable to assume that the percent changes across slices of data like country, device, etc. are exchangeable. Thus, one could build a hierarchical model where the percent change for each slice is modeled with a random effect, allowing for borrowing of information across slices. Another possible extension is to include in the model pre-period data from users not in the experiment but from the same population. A power prior \cite{ibrahim2000power} could be used to control the extent these additional data are used to inform the posterior of the pre-period mean. 

An open source R package implementing the methods described in the paper is freely available at \url{https://google.github.io/abpackage}.

\section*{Acknowledgments}
Many individuals contributed to this project at various stages of its development. Among them I am particularly grateful to David Diez, whose early discussions inspired this work, Wael Salloum for strongly believing in my ability to make this project succeed, and Maria Terres for her many suggestions which resulted in a much clearer and better organized paper. 

\bibliographystyle{apalike}
\bibliography{example} 

\end{document}